\documentclass[twocolumn,
prd,secnumarabic,nobalancelastpage,amsmath,amssymb,superscriptaddress,
nofootinbib]{revtex4-2}

\usepackage{graphicx} 
\usepackage{xcolor}
\newcommand{\req}[1]{Eq.\,(\ref{#1})} 
\newcommand{\rs}[1]{section~\ref{#1}} 
\newcommand{\rf}[1]{Fig.\,\ref{#1}}

\newcommand{\fracg}{\frac{g}{2}}

\begin{document}

\title{Study of QED singular properties for variable gyromagnetic ratio  $g\simeq 2$}

\author{Johann Rafelski}
\author{Stefan Evans}
\author{Lance Labun}
\affiliation{Department of Physics, The University of Arizona, Tucson, Arizona, 85721 USA}

\begin{abstract}Using the external field method, {\it i.e.\/} evaluating the effective action $V_{\mathrm{eff}}$ for an arbitrarily strong constant and homogeneous field, we explore nonperturbative properties of QED allowing arbitrary gyromagnetic ratio $g$. We find a cusp at $g = 2$ in: a) The QED $b_0$-renormalization group coefficient, and in the infinite wavelength limit in b) a subclass containing the pseudoscalar ${\cal P}^{2n}= (\vec E\cdot\vec B)^{2n} $ of light-light scattering coefficients. Properties of $b_0$ imply for certain domains of $g$ asymptotic freedom in an Abelian theory.
\end{abstract}

\maketitle

\section{Introduction}

No known particle has exactly the Dirac value $g_\mathrm{ D}\equiv 2$ of the gyromagnetic ratio $g$. Determination of the higher order vacuum fluctuation correction to $g_\mathrm{D}$ provides the most precise test of perturbative QED (pQED)~\cite{Kinoshita:2005sm}. However, should $g=g_\mathrm{D}$ be a singular point, a pQED perturbative expansion is not appropriate considering that the point-like electron (or muon or tauon) to start with have $g\ne 2$ due to modifications introduced by electromagnetic interactions with particles outside of the QED framework -- these, for example, are EM interactions with quark fluctuations in the vacuum. 

The aims of this work are: To recognize, using the method of second order fermions, a singularity at $g=g_\mathrm{D}$; to study the nature of this singularity; and to lay the foundation for a theoretical framework allowing exploration of the $|g|>g_\mathrm{D}$ domain. We furthermore compare our results with the first order formulation and describe non-renormalizable aspects of this approach. This work develops the insights presented in Ref.\cite{Rafelski:2012ui}.

As we discuss in Ch.~\ref{VarMM} in more detail there are two different approaches to study of $g\ne g_\mathrm{ D}=2$. Aside of the popular first order formulation there is a long established but less well known method involving second order fermions~\cite{Morgan:1995, Espin:2013, Espin:2015bja}. Fock first studied such an equation for the case of particles with $g=2$~\cite{Fock:1937dy}, as a second order (squared) Dirac equation. Feynman, Gell-Mann and Brown subsequently used it to study weak interactions~\cite{Feynman:1958ty, Brown:1958tc}. More recently, Veltman studied the case of particles with $g\neq2$~\cite{Veltman:1997am}, and showed that this path produces a renormalizable field theory, confirming a possible framework for studying particles with anomalous moments arising from nonperturbative interactions, such as would be expected for a lepton from outside of the QED domain. Even more recently, this work has been continued by Araujo, Napsuciale and Martinez~\cite{VaqueraAraujo:2012qa,AngelesMartinez:2011nt}, evaluating the $g$-dependent beta function and self-energy.

To achieve our goals we consider in this work the extension to $g\ne 2$ based on the renormalizable dimension-4 action~\cite{VaqueraAraujo:2012qa,AngelesMartinez:2011nt}. We study the vacuum properties in the presence of external constant and homogeneous electromagnetic fields, integrating out fluctuations of spin-1/2 particles with $g\ne g_\mathrm{ D}$. The resulting effective potential $V_{\mathrm{eff}}$ is a generalization of the Heisenberg-Euler-Schwinger (HES) effective action~\cite{Heisenberg:1935qt,Weisskopf,Schwinger:1951nm,Reuter:1996zm,Dunne:2004nc} to arbitrary value of $g$. The result is regular for all $|g|\le g_\mathrm{ D}$~\cite{Labun:2012jf}.

For $|g|> g_\mathrm{ D}$ the HES proper time effective action becomes non-integrable. Using the Weisskopf~\cite{Weisskopf} Landau level summation method, we obtain an extension of the effective action $V_{\mathrm{eff}}$ to all values $|g|> g_\mathrm{ D}$, which shows that $g=g_\mathrm{ D}$ (and other periodic recurrent values) is a cusp point as a function of~$g$. This extension resolves the known difficulties in the theoretical framework of $g\ne g_\mathrm{ D}$ theories~\cite{Veltman:1997am}. Moreover, considering the beta-function coefficients, we demonstrate the domains for which asymptotic freedom arises as a function of $g$. 

In our approach the self-adjointness requirement helps to select the physical (periodic) Landau eigen energy spectrum. A confirmation of this analytic extension as proposed here was developed applying ideas seen in the work of Nikishov~\cite{Nikishov:1970br} and Kim~\cite{Kim:2009pg} on Bogoliubov coefficient summation, to obtain $V_{\mathrm{eff}}$ with $g\ne g_\mathrm{ D}$ in Ref.~\cite{Evans:2022fsu, Evans:2022EB}.

In Ch.~\ref{periodicg} we present the periodic in $g$ Landau level summation, using the second order fermion formulation. We show how this extends the prior results for effective action, limited to $|g|\leq g_\mathrm{ D}$, Ch.~\ref{KGPgle2}, to $|g|> g_\mathrm{ D}$ in Ch.~\ref{KGPgge2}. We show the cusp arising in the $g$-dependent $\beta$-function in Ch.~\ref{ChBeta}, and the light-light scattering contribution to effective action in Ch.~\ref{Chlightlight}. We compare the second order to the first order fermion formulation of effective action in Ch.~\ref{DPpartA}. While the latter is non-renormalizable, we extract the $g$-dependent properties following the regularization schemes of Dittrich~\cite{Diet78} and Ferrer {\it et. al.\/}~\cite{Ferrer:2015wca}. In summary we discuss further work to address $g-2$ action in localized fields, and possible observable phenomena stemming from nonperturbative $g-2$ behavior in magnetar field strength environments.

\section{Variable magnetic moment $\mathbf{|g|\ne g_\mathrm{ D}}$}
\label{VarMM}
A frequently explored way to account for $ |g|\ne g_\mathrm{ D} $ is via a first order formulation: to complement the Dirac action with an incremental Pauli interaction term $\delta\!\mu\,(\vec \sigma\cdot \vec B+i\vec \alpha \cdot \vec E)=\delta\!\mu\,\sigma_{\alpha\beta}F^{\alpha\beta}/2$. 
$\vec E, \vec B$ are the electromagnetic fields, $F^{\alpha\beta}$ the electromagnetic field strength tensor, $\sigma_{\alpha\beta}=(i/2)[\gamma_{\alpha},\gamma_{\beta}]$ with $\gamma_\alpha$ the usual Dirac matrices, and $\vec \sigma$, and $\vec \alpha=\gamma_5\vec\sigma$ are the Pauli-Dirac matrices. However, such an incremental Pauli interaction is a dimension 5 operator, $[\,\overline\psi\sigma_{\alpha\beta}F^{\alpha\beta}\psi]=L^{-5}$. The coefficient $\delta\!\mu$ consequently has dimension length, which in the case of a composite particle such as the proton, is naturally related to the particle size. Therefore this Dirac-Pauli (DP) modification of the Dirac equation has been a popular tool to describe to the lowest order the magnetic moment dynamics of a composite particle of finite size, for example a proton.

As noted, another approach uses second order fermions adding the full Pauli spin-interaction term to the Klein-Gordon action
\begin{equation}
\label{EoMg}
{\cal L}=\bar\psi\left[\Pi^2-m^2-\frac{g}{2} \frac{e \sigma_{\alpha\beta}F^{\alpha\beta}}{2}\right]\psi,
\end{equation}
where $\Pi_\alpha=i\partial_\alpha+eA_\alpha$. Note that the dimension of the $\psi$ field is $[\psi]=L^{-1}$ and consequently the Pauli interaction is dimension 4. We refer to the study of QED based on~\req{EoMg} as $g$-QED, and the dynamical equation following from~\req{EoMg} as the Klein-Gordon-Pauli (KGP) equation. $g$-QED is the $s=1/2$ case in the study of particles of all spins in the Poincar\'e group framework initiated by Rarita and Schwinger~\cite{Rarita:1941mf, Napsuciale:2006wr}. For related developments, see references in the introduction to Ref.\cite{VaqueraAraujo:2012qa}. For a first discussion of spectroscopic experiments capable of distinguishing between KGP and DP approaches in the near future, see Ref.\,\cite{Steinmetz:2018ryf}.

Since there are at least two distinct paths to introducing $g\ne 2$ corrections into relativistic particle dynamics, the questions are, in what sense these could be equivalent, and if not, which of the two forms is appropriate for the study of particle dynamics and/or vacuum structure and under what conditions:\\[0.1cm]
1) The DP approach, involving a di\-men\-sion-5 operator, requires new counter terms in each order, rendering it non-renormalizable~\cite{Veltman:1997am}. This, in our opinion, limits the DP approach to situations in which the physical particle properties are known and vacuum fluctuations need not be considered. Even so, we see in literature the DP method applied to both vacuum fluctuation and the effective action evaluation in QED, see for example Refs.~\cite{Diet78, Ferrer:2015wca, Ferrer:2019xlr, Adorno:2021xvj}. \\[0.1cm]
2) In $g$-QED the magnetic moment remains point-like; $g\neq g_{\rm D}$ does not require a higher dimensioned operator. Therefore the quantum field theory requires a finite number of counter terms and is renormalizable~\cite{VaqueraAraujo:2012qa,AngelesMartinez:2011nt}; vacuum fluctuations can be considered in any perturbative order. \\[0.1cm]
3) It should be remembered that in $g$-QED an expansion around $g=2$ requires additional consideration since the natural expansion occurs around $g=0$. Properties of the KGP-originating non-perturbative effective action were considered for general spin in Ref.~\cite{Kruglov:2001dp}. However, this work did not recognize for spin-1/2 the restricted to $-2\le g\le 2$ validity domain of the perturbative approach, which arises due to convergence properties of the proper time Schwinger integral. \\[0.1cm]
4) Another study of quantum field amplitudes with an anomalous moment~\cite{Larkoski:2010am} also arrives at a second-order effective theory, but for a reduced two-component spinor. Given the derivation and properties of their effective theory, we believe that an exact relation between KGP and DP approaches can at best arise in an infinite order resummation in some specific applications.

Veltman has considered reduction of the number of dynamical components working in a two-component formulation. However, there are unresolved challenges~\cite{Veltman:1997am} in particular related to self-adjointness of the resulting spectrum and thus conservation of probability in temporal evolution. By individually characterizing states comprising the physical KGP Landau eigen energy spectrum, we will present another resolution of this problem that works in presence of externally applied fields.

Considering that~\req{EoMg} is 2nd order in time and has four components, the number of dynamical degrees of freedom present in~\req{EoMg} is 8. That is, there are twice as many degrees of freedom as in usual Dirac theory. For the case $g=2$,~\req{EoMg} can be presented as the square of the operator $\gamma_5 D,\ D=\gamma_\alpha(i\partial_\alpha+eA^\alpha)-m$ and $\gamma_5=i\gamma^0\gamma^1\gamma^2\gamma^3, \gamma_5^2=1$ is the 5th Dirac matrix. This means that for $g=2$,~\req{EoMg} comprises exact duplication of the Dirac degrees of freedom, the second set with opposite sign of mass $m$ which means with opposite association of the sign between magnetic moment and electric charge, see paragraphs below; for $g\ne 2$ one must search for a projection restricting the full Hilbert space to the physical states.

Considering these features of the two formulations, for the purpose of deriving a renormalizable $g\ne 2$ extension of the QED effective action, the second order KGP expression is theoretically favored. We begin with the KGP formulation and then present the DP case for comparison.

\section{Eigenvalue-sum periodicity as a function of $g$}
\label{periodicg}

We seek to identify the physics content of the 8 degrees of freedom and to separate the Hilbert space into two equal size parts that each individually comprises a complete set of states at a fixed given value of $g$. To do so, we consider the Landau-orbit spectrum of the operator in brackets in~\req{EoMg} in the presence of a constant magnetic field $\vec B$
\begin{equation}
\label{spectrum}
E_n =\pm\sqrt{m^2+p_z^2+Q|e\vec B|\,[(2n\!+\!1) \mp g/2]},\quad Q=\pm 1,
\end{equation}
where $p_z$ is the one dimensional continuous momentum eigenvalue and $n$ is the Landau orbit quantum number. 

We have made explicit the presence of 8 eigenvalues for each value of $\vec B$, corresponding to all different possible choices of the three $\pm$-sign sets. There are the usual two roots in~\req{spectrum}, a known feature of relativistic dynamics also seen in the Landau spectrum of the Dirac equation where the negative energy states become positive energy antiparticle `hole' states. There is a new spectrum duplication related to two possible values of $Q$. This factor arises from two possible particle spin projections onto magnetic field, corresponding to the spin degeneracy.

To see how $Q$ can be restricted, let us consider~\req{spectrum} in the form
\begin{equation}\label{spectrum1}
K=\frac{E_n^2 -m^2-p_z^2}{ |e\vec B|}=Q\,[(2n\!+\!1) \mp g/2],\quad Q=\pm 1.
\end{equation}
The quantity $K$ is shown in the top portion of~\rf{fig:Veffexpg} as a function of $g$. We see that between $-2\le g\le 2$, there is an exact duplication of the spectrum corresponding to $Q=1$ and $Q= -1$. These are two sectors of the Hilbert space with the same physical content. The `squared' Dirac operator produces two eigenstate-space copies which can be separated in particular applications. Without restriction of generality the $Q=-1$ eigenvalues can be therefore omitted. Thus for $-2\le g\le 2$ the effective action is obtained by the usual procedure, and the results have already been presented~\cite{Labun:2012jf}.

\begin{figure}[t]
\begin{center}
\includegraphics[width=0.99\columnwidth]{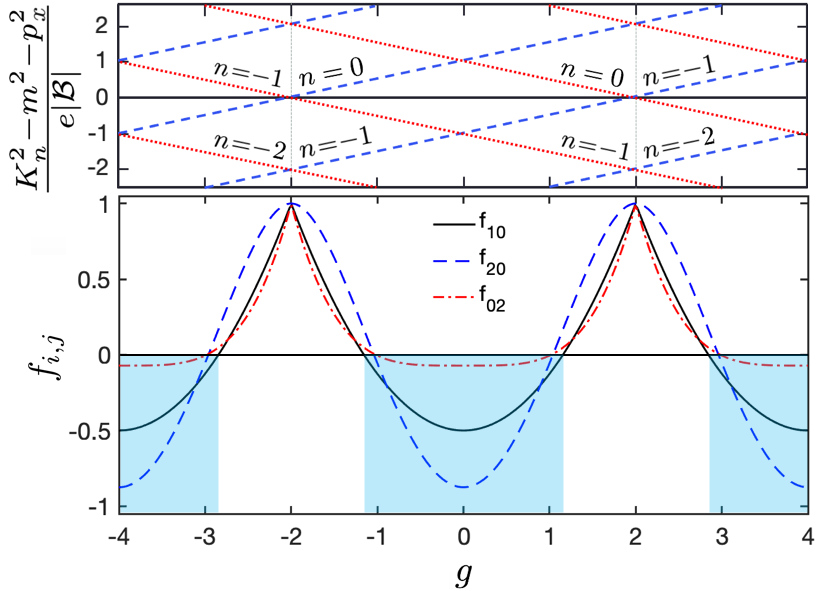}
\caption{Top: {Squared} eigenvalues~\req{spectrum} of KGP in magnetic field; the positive and negative valued domains separated by the solid line are for $Q=+1$ and $Q=-1$ respectively. Bottom: coefficient functions: $f_{1,0}(g)$ as defined in~\req{betafunc} and $f_{0,2}$ and $f_{2,0}$ as defined in~\req{VeffexpB} and~\req{VeffexpC}. Two full periods are shown. The asymptotically free domains of $g$ where functions $f_{i,j}$ change sign are shaded. \label{fig:Veffexpg}}
\end{center}
\end{figure}

For $|g|>2$, new and non-perturbative physics content arises for external fields of any strength, including arbitrarily weak. First we note that taking~\req{spectrum} expression at face value, naively some eigenstates could have $E^2<m^2$, which for strong magnetic fields $|e\vec B|\gg m^2$ acquire complex eigenvalues. In a magnetic environment, such states cannot be admitted in the spectrum. This situation differs from the $m^2+p_z^2\to 0$ limit, in which states having $K<0$ signal instabilities of the conventional vacuum state~\cite{Savvidy:1977as,Nielsen:1978rm}.

To compute the effective action we must define which states contribute to the physical spectral sum. The first step is to accomplish (like for the case $|g|\le 2$) separation of the Hilbert space into two sectors. We divide the states according to whether $K\geq 0$ or $K\leq 0$ and denote the respective sectors ${\cal K}^{\pm}$. The limit $K=0$ where two states coincide occurs at $g=2$, since the KGP operator can be written as the exact square of the Dirac operator. This situation recurs with the shift of $g$ by $4k, k\in \mathbb{Z}$. There is no change in the number of states in each of the Hilbert space sectors ${\cal K}^{\pm}$ as an equal number of single particle states is exchanged between both sectors.

The principle we use to determine which states enter the spectral sum is that there should be no states with complex eigenvalues in a constant magnetic field. In the notation just introduced, we require $K\geq 0$ and the ${\cal K}^+$ sector be chosen as representing the physical spectrum. This is an extension from the regular case $|g|\leq 2$, where the usual procedure sums over the $Q=+1$ states and is equivalent to summing over the ${\cal K}^+$ state space. As $K\geq 0$ implies $E^2\geq m^2$, the physics is a continuous extension of the case $g=2$, for which it is proved that $E^2\geq m^2$ for arbitrary magnetic fields, i.e. there are no bound states~\cite{Gornicki:1987hv}. 

Looking far outside the principal domain $-2\le g\le 2$, we see that relativistic Landau eigenstates cross between ${\cal K}^\pm$ at each $g_k=2+ 4k, k\in \mathbb{Z}$. As the graphic representation top frame of~\rf{fig:Veffexpg} shows, for each of the Hilbert space sectors ${\cal K}^{\pm}$ we have periodicity of the Landau levels a function of $g$. Therefore, the sum $\sum_{n}E_n$ over ${\cal K}^+$ leading to the real part of $V_{\mathrm{eff}}(\vec B^{\,2})$ is a periodic function of $g$, a result we will find explicitly. This periodicity does not apply to individual Landau eigenvalues as is seen in~\req{spectrum}. In computation of vacuum fluctuations the truncation of the Landau eigenstate $n$-sum to any finite value breaks the periodicity as well.

The choice of ${\cal K}^+$ as the physical state space has clear advantages and resolves the challenges encountered by Veltman~\cite{Veltman:1997am}: In addition to maintaining self-adjointness of the KGP system, it makes the quantum field theories based on semi-spaces ${\cal K}^{\pm}$ each individually unitary, because the number of states is conserved in transiting through the singular points e.g at $|g|=2$, and for $|g|>2$ we omit the nonphysical solutions. Moreover, our proposal resolves the spectrum and by extension 
its summation comprising the effective action $V_{\mathrm{eff}}$ for all $g$ including the domain $|g|>2$. Had we separated the sectors along the sign of $Q$, the contents of the theory would be different for $|g|>2$ and unitarity would be violated since the `wrong' levels would be included in the physical half-space. An independent confirmation of the choices made is accomplished via the Bogoliubov coefficient summation method, a point we return to in~\rs{Bogsum}.

\section{Effective action for $\mathbf{|g|\le 2}$} 
\label{KGPgle2}

We briefly summarize results for ${|g|\le 2}$~\cite{Labun:2012jf}, as these are needed to understand the novel case of ${|g|>2}$. For constant fields the effective action is manifestly covariant and can be written as a function of the Lorentz-invariant field-like quantities $a,b$
\begin{align}
\label{Pdef}
&b^2-a^2= \vec B^2-\vec E^2 = \frac{1}{2}F_{\alpha\beta}F^{\alpha\beta}\equiv 2{\cal S}
\;, 
\\ \nonumber
&(ab)^2 = (\vec E\cdot\vec B)^2 =\left(\frac{1}{8}F^{\alpha\beta}\varepsilon_{\alpha\beta\kappa\lambda}F^{\kappa\lambda}\right)^2 \equiv {\cal P}^2 
\;,
\end{align}
where $\pm a$ are electric-field-like and $\pm ib$ are the magnetic-field-like eigenvalues of $F^{\alpha\beta}$. $a$ is considered electric-like because $a\to|\vec E|$ on taking the limit $b\to 0$, and similarly $b\to |\vec B|$ in the limit $a\to 0$.

The Schwinger-Fock proper time method~\cite{Schwinger:1951nm} to evaluate the effective action exploits properties of the \lq squared\rq\ Dirac equation, and thus it can be used to study the arbitrary value of $g$. The effective action can be written in the form
\begin{equation} 
V_{\mathrm{eff}} = \frac{1}{8\pi^2}\int_{0}^{\infty}\frac{du}{u^{3}}\,e^{-i(m^2-i\epsilon)u} F(eau,ebu,\frac{g}{2}).
\label{Veffab}
\end{equation}
For $g=0,2$, the proper time integrand $F (eau,ebu,g)$ was reviewed in Ref.~\cite{Dunne:2004nc}. The generalization throughout the interval $|g|\leq 2$ is accomplished by inserting into Schwinger's Eq.\,(2.33) in the last term a co-factor $g/2$ leading to~\cite{Labun:2012jf}.
\begin{equation} 
\label{VeffabF} 
F(x,y,\frac{g}{2})=\frac{x\cosh(\frac{g}{2} x)}{\sinh x}
\frac{y\cos(\frac{g}{2} y)}{\sin y}-1
,\ \ \left|\frac{g}{2}\right|\le 1 .
\end{equation}
The subtraction $-1$ in~\req{VeffabF} removes the field-independent constant. The logarithmically divergent charge renormalization term is isolated and discussed below. Note that~\req{Veffab} would be divergent for $ |g|> 2$ if~\req{VeffabF} were to be used in this domain.

\section{Effective action for $\mathbf{|g|>2}$} 
\label{KGPgge2}

\subsection{Evaluation based on periodic Landau level spectrum}

To extend~\req{VeffabF} to $|g|>2$, we consider in more detail the eigenvalue summation method we introduced above, following the work of Heisenberg and Euler~\cite{Heisenberg:1935qt} and Weisskopf~\cite{Weisskopf}. The mathematical tool used was the L.\,Euler summation formula, leading to the Bernoulli functions $B_{2k}(x)$ and Bernoulli numbers ${\cal B}_{2k}\equiv B_{2k}(0)$. The sum of the Landau energies~\req{spectrum} involves the form $\sum_n f(x+n)$. L.\,Euler developed the technique for such sums, which manifest an integer shift symmetry in the variable $x\to x+n'$~\cite{Euler,Apostol}. Due to this shift symmetry, the Bernoulli functions $B_{2k}(x)$ that arise in the context of L.\,Euler summation of Landau energies $E_n$,~\req{spectrum} are {\it periodic}, given by the Fourier series~\cite{Luo} 
\begin{equation}\label{Bernoullifns}
\tilde B_{2k}(t)=(-1)^{k-1}\frac{(2k)!}{2^{2k-1}}\sum_{n=1}^{\infty}\frac{\cos(2\pi nt)}{(n\pi)^{2k}},
\end{equation}
(here only needed for an even value of index, $2k$). In the unit interval, $0\leq t\leq 1$, the periodic Bernoulli functions are equal to the Bernoulli polynomials, e.g. $\tilde B_{2}(t)=B_2(t)=t^2-t+1/6,\ 0\leq t\leq 1$. Outside the unit interval, the periodic Bernoulli functions~\req{Bernoullifns} $\tilde B_{2k}(t)$ repeat the polynomials' behavior on $0\leq t\leq 1$ in each subsequent period.

Dividing the Landau energies by $2|e\vec B|$ to make the coefficient of $n$ unity, we see that $t\to g/4+1/2$ and hence we recognize that the periodic Bernoulli functions with argument $t=g/4+1/2$ appears in the effective action, arising from the summation of eigenvalues. The explicit representation of the argument of~\req{Veffab} in terms of Bernoulli functions is arrived at employing the analytic transformation of the integrand of~\req{Veffab}~\cite{Muller:1977mm,Cho:2000ei}.
%
\begin{align}\label{meroexpB} 
&\;F(x,y,\frac{g}{2})
=(x^2\!-\!y^2)\,2\!\sum_{n=1}^{\infty}\frac{\cos n\pi(\frac{g}{2}+1)}{(n\pi)^2} 
\\ \nonumber
&\;
+2\sum_{n=1}^\infty\frac{(-1)^n\cos(\frac{g}{2} n\pi)}{n^2\pi^2}
\Big(\frac{y^4}{y^2-n^2\pi^2} - \frac{x^4}{x^2+n^2\pi^2}\Big)
\\ \nonumber 
&\; 
+ 4x^2y^2\!\!\sum_{n,\ell =1}^\infty
\frac{(-1)^{n+\ell }\cos(\frac{g}{2} n\pi)\cos(\frac{g}{2} \ell \pi)}{(y^2-n^2\pi^2)(x^2+\ell ^2\pi^2)}
\;.
\end{align}
 
In~\req{meroexpB} to assure the necessary periodicity we have introduced, in accordance with~\req{Bernoullifns}, a series of Bernoulli functions with $t=g/4+1/2$. Equation\,(\ref{meroexpB}) agrees exactly with the known expansion~\cite{Cho:2000ei} in the domain of~\req{VeffabF} $|g|\leq 2$ and provides an analytical continuation into the domain $|g|>2$ having the periodicity property of the effective action identified in study of the full set of eigenvalues. Even after the removal of the charge renormalization subtraction term (first term on RHS of~\req{meroexpB}) the finite remainder of the effective action is manifestly periodic in $g$.

Upon performing the proper time integral~\req{Veffab}, each term in~\req{meroexpB} produces a well-defined result for all $g$. The form~\req{meroexpB} is thus an analytic and convergent extension to $|g|>2$ developed using the Euler summation of the eigenvalues~\req{spectrum}. We note that~\req{meroexpB} extends the pure magnetic action based on Weisskopf summation case to arbitrary EM fields via analytical continuation. An independent verification of this continuation was obtained for pure electric fields~\cite{Evans:2022fsu}, and is addressed for field configurations with nonvanishing pseudoscalar $(\vec E\cdot\vec B)$ in Ref.\,\cite{Evans:2022EB}. This then completes the proof that the here proposed analytical continuation is unique.

\subsection{Validation with Bogoliubov coefficient summation} 
\label{Bogsum}

We check validity of the Landau level summation with an alternate derivation method, based on the Bogoliubov coefficient summation to compute the imaginary part of effective action as a function of $g$, for a pure electric Sauter potential step. We briefly summarize the result presented in~\cite{Evans:2022fsu}, which evaluates the sum of the tunneling probabilities for electron-positron pairs to materialize. 

Since the asymptotic states of the KGP wave equation in a localized potential step are well defined, we can compute the Bogoliubov coefficient as the ratio $\mu=A/B$ between the incident $A$ and reflection coefficients $B$. Summing the absolute value of the coefficients gives, integrating over states capable of tunneling and summing spin projections, the imaginary part of effective action
\begin{align}
\label{SumB}
\mathrm{Im}[V_{\mathrm{eff}}]=-\int \frac{dE dp^2_\perp}{(2\pi)^3}\sum_{\pm}\ln|\mu|
\;.
\end{align}

Since the summation~\req{SumB} spans a finite range of continuous states determined by the height of the Sauter step, all the states comprise the physical spectrum. This bypasses the need for identification of the physical spectrum as in~\rs{periodicg}, where the infinitely spanning constant field idealization necessitates self-adjoint extension. The summation~\req{SumB} produces the same periodic spectrum where the resulting effective action repeats with each shift in $g$ by $4k, k\in \mathbb{Z}$, Fig.~(1) of~\cite{Evans:2022fsu}. The electric and magnetic dominated cases can thus be transformed from one to another via $E\leftrightarrow iB$.

\section{Nonperturbative in $\mathbf{g}$ renormalization group $\mathbf{\beta}$ function} 
\label{ChBeta}

The first non constant term on the right hand side of~\req{meroexpB} proportional to $a^2-b^2$ isolates the logarithmically divergent one-loop $\mathcal{O}(\alpha)$ $V_{\mathrm{eff}}$ subtraction required for charge renormalization. The coefficient of this term is related to the $\beta$-function coefficient $b_0$ as is discussed e.g. in section 5.1 in Ref.\,\cite{Dunne:2004nc}. 

We now evaluate the running of the coupling constant $\alpha$ within the $g$-QED loop expansion of the $\mathbf{\beta}$-renormalization function 
\begin{equation}\label{betaf}
\beta \equiv \mu\frac{\partial\alpha}{\partial\mu}, \quad 
\beta(\alpha)=-\frac{b_0}{2\pi}\alpha^2 +\frac{b_1}{8\pi^2}\alpha^3+\ldots \,.
\end{equation}
The first sum in~\req{meroexpB}, for $g=2$, $\sum_{n=1}^\infty 1/(\pi n)^2=1/6$ and implies the value of $b_0=-4/3$, where factor 4 indicates the 4 components of spin-1/2 particle. For arbitrary $g$, $b_0(g)$ is obtained using~\req{Bernoullifns} to identify this sum as $\tilde B_2(g/4+1/2)$. The character of this function is manifested by reconnecting periodic domains of the familiar Bernoulli polynomial $B_2(t)=t^2-t+1/6$ and the resulting $b_0(g)$ coefficient is given in each domain $g\in [g_{k-1},g_k]$
\begin{equation}
 b_0=\:-\frac{4}{3}f_{1,0}(g)
=-\frac{4}{3}\left(\frac{3}{8}(g-4k)^2-\frac{1}{2}\right), \label{betafunc}
\end{equation} 
where $f_{1,0}(g)$ is shown in bottom frame of~\rf{fig:Veffexpg}. The subscripts of $f_{i,j}$ indicate the powers of the Lorentz invariants in polynomial expansion $f_{i,j}{\cal S}^i {\cal P}^j$ in~\req{meroexpB}. We see in~\rf{fig:Veffexpg} that as a function of $g$, the Dirac value $g_{\rm D}=\pm 2$ is an upper cusp point with $f_{1,0}(g)\leq f_{1,0}(2)=1$. For clarity, two periods are shown in~\rf{fig:Veffexpg}. 

Note that our result arises in $g$-QED, applying a nonperturbative method in $g$ to one loop expansion. This approach is necessary in order to obtain the behavior of the $\beta$-function for $|g|>2$. At $g=\pm 2$ we find the unexpected cusp. This feature is missing in perturbative consideration of $\beta(g)$ at one loop level which produces the same functional dependence on $g$ as seen in~\req{betafunc} setting $k=0$. As our study shows, a perturbative expansion around $g=0$ in $g$-QED has a finite convergence interval $|g|\le 2$. This was also seen in the Schwinger proper time integral of the effective action.

The following properties of the renormalization group coefficient $b_0(g)$ shown in~\rf{fig:Veffexpg} are noteworthy:\\
1.) The pQED expands around $g=\pm 2$, which points are identified as being non-analytic at $g_\mathrm{D }=2$ in the $g$-QED framework.\\
2.) For any value of $g$ not at the cusp $g_\mathrm{D }=2$, the magnitude $|b_0|$ compared to its value at $g=2$ decreases, and thus the speed of `running' decreases. Considering that the coefficient of the magnetic spin term in~\req{EoMg} is dimensionless, no new scale appears in association with $g$. \\
3.) The presence of the cusp in $b_0$ implies that the running coupling of $g$-QED, comprises the cusp as well. \\
4.) A cross check and confirmation of our result for $b_0(g)$ is obtained in perturbative domain considering the limit $g\to 0$ where $b_0(g\to 0)$ differs as expected in sign and the number of degrees of freedom from the known behavior of scalar particle `QED'. 
\\
5.) In the principal domain $|g|\le 2$, the functional dependence on $g$ we find agrees with the result Eqs.\,(53--57) seen in Ref.\,\cite{AngelesMartinez:2011nt}. Specifically, the leading term for large $q^2$ of the vacuum polarization function, evaluated within the framework of $g$-QED is $-\alpha b_0(g)/(2\pi) \ln (-q^2/m^2)$, seen explicitly in Eq.\,(55) of Ref.\,\cite{AngelesMartinez:2011nt}.\\ 
6.) As the above limit shows, for a range of appropriate gyromagnetic moment values $g$ (including $g=0$), $b_0(g)>0$ is {\em possible}. This produces asymptotic freedom behavior for Abelian fermions. The switch between the infrared stable and the asymptotically free behavior occurs in the principal $g$-domain twice, at $g =\pm 2/\sqrt{3}=\pm 1.155$ and continues periodically e.g. for $g=4-2/\sqrt{3}=2.845$. This mechanism of asymptotic freedom generation by $g$-driven sign reversal is implicit in Eq.\,(56) of Ref.\,\cite{AngelesMartinez:2011nt} (valid in principal domain $|g|\le 2$), but the new mechanism allowing Abelian confinement has not been recognized there. The values of $g$ where the sign of the functions $f_{i,j}$ changes is indicated in~\rf{fig:Veffexpg}, up to periodic recurrence, agreeing with the periodic result based on Bogoliubov coefficient summation reported in~\cite{Evans:2022fsu}.

\section{Light-light scattering as function of $\mathbf{g}$} 
\label{Chlightlight}

We find that the cusp at $|g|=2$ reappears in the Heisenberg-Euler action, in the light by light scattering. For the general case of both electric and magnetic fields present, using~\req{meroexpB} we find up to fourth order in the fields
\begin{subequations}
\begin{align}
\label{Veffexp} 
V_{\mathrm{eff}} \simeq &
\frac{\alpha}{2\pi}\frac{e^2}{45m^4}
\!\left(4f_{2,0}\:{\cal S}^2+7f_{0,2}\:{\cal P}^2\right) 
\;,
\\ 
\label{VeffexpB} 
f_{2,0}(g)
&=-30\tilde B_4(g/4+1/2)
\\ \nonumber
&=-\frac{15(g-4k)^4}{128}
+\frac{15(g-4k)^2}{16}
-\frac{7}{8} 
\;,
\\ 
\label{VeffexpC} 
f_{0,2}(g)
&=-\frac{60}{7}\left[\tilde B_4\left(\frac g 4 +\frac 1 2\right)
 -3\tilde B_2^2\left(\frac g 4 + \frac 1 2 \right)\right]
\\ \nonumber
&=\frac{15(g-4k)^4}{224}-\frac{1}{14} 
\;,
\end{align}
\end{subequations}
where both $f_{2,0}$ and $f_{0,2}$ are normalized to $g=2$ values and presented in~\rf{fig:Veffexpg}. $f_{0,2}$ includes a product of two Bernoulli functions with cusp and so has a steeper cusp. In general, our finding is that all $f_{i,j}(g)$ for $j>0$ have cusps at $g=2$, whereas all $f_{i,0}(g), i>1$ are continuous and differentiable at $g=2$, being proportional to higher order $>2$ Bernoulli functions that have vanishing derivatives at $g=2$. Said in plain English: only coefficients of terms involving powers of the pseudo-scalar field invariant ${\cal P}^2=(\vec E\cdot\vec B)^2$ display cusps at $g=2$.

\section{Comparing the Klein-Gordon-Pauli and Dirac-Pauli approaches}
\label{DPpartA}

Thus far we have discussed the effective action generated in the KGP approach. We now briefly review the DP expression and past efforts to use it to evaluate effective action. The DP equation complements the Dirac equation with the Pauli term describing the anomalous magnetic moment interaction:
\begin{align}
\label{DPgeneral} 
{\cal L}_{\mathrm{DP}}=\bar\psi\Big(\gamma^\mu\Pi_\mu-m -
\Big(\frac g2-1\Big)\frac{e}{2m} \frac{\sigma^{\mu\nu}F_{\mu\nu}}2 \Big)\psi
\;.
\end{align}
This form is only equal to the KGP expression~\req{EoMg} when $g=2$. While solving~\req{DPgeneral} is significantly more challenging than it is for the ($g=2$) Dirac equation, there is a class of EM field configurations for which exact DP solutions exist. Notable solutions include the case of electrons in a constant magnetic field~\cite{Tsai:1971zma}, and more recently charge-neutral particles in a Sauter pulsed magnetic field~\cite{Adorno:2021xvj}. 

Below we review and develop further prior efforts to evaluate the effective action $V_\mathrm{eff}^\mathrm{DP}$ in the DP context showing the emergence of novel properties similar to those we found in the KGP $g\neq2$ case. The main challenge is how to handle the non-renormalizability~\cite{Veltman:1997am}:
\begin{enumerate}
\item Dittrich derived the DP based $V_\mathrm{eff}^\mathrm{DP}$ action using the Schwinger proper time method~\cite{Diet78}. This work squares the DP equation and omits a certain cross term, which reproduces the KGP equation but with a modified mass term, and obtains a renormalizable effective action. 
\item Ferrer {\it et. al.\/}~\cite{Ferrer:2015wca} several decades later derived the DP $V_\mathrm{eff}^\mathrm{DP}$ action using the Weisskopf summation method~\cite{Weisskopf}. This work takes a different approach using the exact DP wave equation, and truncating the anomalous contributions in the Landau level sum to obtain a finite result for $V_\mathrm{eff}^\mathrm{DP}$. To our knowledge this is the only evaluation of the real part of $V_\mathrm{eff}^\mathrm{DP}$ based on exact solutions to the DP wave equation. 
\item We extend the work of Ferrer to higher order Landau levels and show explicitly how the DP action is linearly divergent when summing Landau level corrections to all orders. Despite this nonrenormalizablility of fully summed DP based action, the truncation by Ferrer allows for a finite action result which we study as a function of $g$.
\end{enumerate}

\subsection{Dittrich approach to the Dirac-Pauli action}
\label{DPpartB}

Dittrich using the Schwinger proper time method derived the DP based $V_\mathrm{eff}$ action~\cite{Diet78}. However, Dittrich formed an incomplete square of the DP operator, by omitting a cross term between the derivative and the Pauli operators. Had this term been kept, it would not be possible to formulate a renormalizable expression with the DP proper time evolution operator. Nonetheless this method may still be a viable approximations allowing for regularization of the otherwise non-renormalizable action. 

We trace the steps of Dittrich's work~\cite{Diet78}: inserting the negative mass counterpart,
\begin{align}
\label{ansatzDP}
\psi=\Big(\gamma^\mu\Pi_\mu+m +
\Big(\frac g2-1\Big)\frac{e}{2m} \frac{\sigma^{\mu\nu}F_{\mu\nu}}2 \Big)\Psi
\;,
\end{align}
gives the second order equation
\begin{align}
\label{DPgeneral2}
&\Big(\Pi^2-\tilde m^2-\fracg\frac e2\sigma^{\mu\nu}F_{\mu\nu}
\nonumber \\ 
&
\;+\Big(\frac g2-1\Big)\frac{e}{4m} \Pi_\beta F_{\mu\nu}\Big[\gamma^\beta,\sigma^{\mu\nu}\Big]\Big)\Psi=0
\;,
\end{align}
where the squared Pauli term $\sigma F$ modifies the mass as~\cite{Diet78}
\begin{align}
\label{tildeM}
\tilde m^2=m^2-\Big(\frac g2-1\Big)^2\frac{e^2}{4m^2}(\vec E\,^2-\vec B\,^2)
\;.
\end{align}
There is also the commutator in~\req{DPgeneral2}: Using the relation $[\gamma^\beta,\sigma^{\mu\nu}]=2i(\eta^{\beta\mu}\gamma^\nu-\eta^{\beta\nu}\gamma^\mu)$, where the Minkowski metric $\eta_{\mu\nu}=\mathrm{diag}(1,-1,-1,-1)$, this term can be written as
\begin{align}
\label{commutDP}
\Pi_\beta F_{\mu\nu}\Big[\gamma^\beta,\sigma^{\mu\nu}\Big]
=4i F_{\mu\nu}\Pi^{\mu}\gamma^\nu
\;,
\end{align}
responsible for mixing between the Landau quantum numbers and continuum momentum dependence~\cite{Tsai:1971zma, Steinmetz:2018ryf}.

To evaluate the effective action, Dittrich omitted the commutator term~\req{commutDP} from~\req{DPgeneral2}. A reason for omitting this term was not given in~\cite{Diet78}. However, the outcome is a viable approximate form of the \lq squared\rq\ DP wave equation 
\begin{align}
&(H_{\mathrm{DP}}-m^2)\Psi=0
\;,
\end{align}
where
\begin{align}
\label{Hdp}
H_{\mathrm{DP}}-m^2=
\Pi^2-\fracg\frac e2\sigma_{\mu\nu}F^{\mu\nu}-\tilde m^2
\;.
\end{align} 
By using~\req{Hdp}, Dittrich was able to employ the proper time approach and obtain a renormalizable action.

Applying $H_{\mathrm{DP}}$ from~\req{Hdp} as the proper time evolution operator, the resulting effective action
\begin{align}
\label{3.22DP0}
V_\mathrm{eff}^\mathrm{DP}=&\,
\frac{i}{2}\int_0^\infty\frac{du}{u}e^{-im^2u}\mathrm{tr}\left<x\right|e^{-iH_\mathrm{DP} u}\left|x\right>
\\ \nonumber 
=&\;
\frac{1}{32\pi^2}\int_0^\infty\frac{du}{u^3}e^{-i\tilde m^2u}
\frac{e^2u^2ab\; \mathrm{tr}e^{iu(g/2)e\sigma F/2}}{\sinh(eau)\sin(ebu)}
\;,
\end{align}
where $V_\mathrm{eff}^\mathrm{DP}$ denotes the DP approach to $g$-dependent action. Carrying out the trace in~\req{3.22DP0} gives the renormalizable expression
\begin{align}
\label{3.22DP}
&V_\mathrm{eff}^\mathrm{DP}=\frac{1}{8\pi^2}\int_0^\infty\frac{du}{u^3}e^{-i(\tilde m^2-i\epsilon)u}
F(eau,ebu,\fracg)
\;,
\end{align}
where $F(eau,ebu,\fracg)$ in~\req{3.22DP} is the same as in the KGP case~\req{VeffabF}. 
While Dittrich originally evaluated the action for pure magnetic fields, this proper time approach leading to \req{3.22DP} allows for both $E$ and $B$ fields to be nonzero. Setting $a\to0$ ($E\to0$) we recover the pure magnetic result Eq.\,(2.9) of~\cite{Diet78}.

Akin to the KGP proper time approach discussed in~\rs{KGPgle2}, here~\req{3.22DP} follows from application of $g\ne2$ directly into the second order proper time evolution operator as a correction to the Schwinger ($g=2$) proper time Hamiltonian. This approach is not guaranteed to be consistent for all $g$, as we have seen in the KGP result in~\rs{KGPgle2}, which is valid only for $|g|\leq2$.

Comparing~\req{3.22DP} to the KGP based action~\req{Veffab}, the only difference is that mass $\tilde m^2$ is used here in place of $m^2$ in the KGP case. Since this difference in mass (\req{tildeM}) is analytic in $E,B$ and $g$, the (non-analytical) singular properties remain, and the Dittrich effective action inherits the singular in $g$ structure, with a cusp at $g=2$. Moreover, setting $|\vec E|=|\vec B|$, the mass $\tilde m^2=m^2$ according to~\req{tildeM}, and the effective action~\req{3.22DP} is identical to~\req{Veffab}, valid in the domain $|g|\le2$. Consequently our evaluation of effective action for $g>2$ in~\rs{KGPgge2} via the periodic Landau eigen energy spectrum (or the independently obtained Bogoliubov summation in~\cite{Evans:2022fsu}) resolves both the KGP and Dittrich\rq s second order DP action in the previously inaccessible domain $|g|>2$.

\subsection{Ferrer evaluation of the Dirac-Pauli action}
\label{DPpartC}
We revisit here the DP action work by Ferrer {\it et. al.\/}~\cite{Ferrer:2015wca}. Their one-loop self-energy correction to the electron propagator builds on the work of Ritus~\cite{Ritus:1972ky}. Two of their key results are: evaluation of the self-energy in the strong magnetic field limit, and implementation of field-dependent $g-2$ corrections into the DP-based Weisskopf Landau level summation as corrections to the HES action.

We recall the zero field limit of the electron $g$-factor to one-loop order
\begin{align}
\label{g0form}
\lim_{|e\vec B|/m^2\to0}
\Big(\frac g2-1\Big)=\frac{\alpha}{2\pi}
\;.
\end{align} 
When strong magnetic fields are switched on, $g$ becomes dependent on both $B$ and the Landau level $n$: $g\to g_n(B)$. For the first Landau level ($n=1$)~Eq.\,(15) of~\cite{Ferrer:2015wca} reads
\begin{align}
\label{T1form}
\lim_{|e\vec B|/m^2\to\infty}
\Big(\frac {g_1}2-1\Big)=\frac{\alpha}{2\pi}\frac{m^2}{2|e\vec B|}\ln^2\Big(\frac{m^2}{2|e\vec B|}\Big)
\;.
\end{align}
For $n>1$ states see~Eq.\,(16) of~\cite{Ferrer:2015wca}. We note that Ferrer {\it et. al.\/} denote the anomaly using a symbol $\mathcal{T}$ such that $g_n(B)/2-1\equiv\frac{2m}{|e\vec B|}\mathcal{T}_n(B)$.

The anomalous moment correction is inserted in the DP Landau orbit spectrum according to Eq.\,(23) of~\cite{Ferrer:2015wca}:
\begin{align}
\label{spectrumDP}
&E_n^{\mathrm{DP}} \!=
\pm\sqrt{p_z^2+\!\bigg(\!\sqrt{2|e\vec B|n+m^2} \pm 
\frac{|e\vec B|}{2m}\Big(\frac {g_n}2-1\Big)\!
\bigg)^2},
\nonumber \\ 
&n\ge1
\;,
\end{align}
compare to the KGP spectrum~\req{spectrum}. \req{spectrumDP}, derived in~\cite{Tsai:1971zma} (see also section~34 of~\cite{Bagrov:1990xp}), is the exact eigen energy obtained by solving the DP wave equation. Comparing the DP wave equation Eq.\,(2) of~\cite{Tsai:1971zma} to~\req{Hdp} in~\rs{DPpartB}, we see that the term omitted by Dittrich is kept here in derivation of~\req{spectrumDP}. Retaining this term leads to the cross term in the parenthesis in~\req{spectrumDP}, responsible for further mixing between quantum number $n$ and $g-2$.

To evaluate the DP effective action, Ferrer {\it et. al.\/} apply the Weisskopf Landau level summation to the spectrum~\req{spectrumDP}. The full summation over Landau levels $n$, and $\pm$-spin projections defines the effective action
\begin{align}
\label{LDPsum0}
V_\mathrm{eff}^\mathrm{DP}
=&\;\frac{|e\vec B|}{8\pi^2}\int_{-\infty}^\infty dp_z \sum_\pm \sum_{n=0}^\infty
|E_n^{\mathrm{DP}}|
\;,
\end{align}
However, Ferrer {\it et. al.\/} approach truncates the AMM dependent contributions to the Landau levels by introducing an upper limit as shown below
\begin{align}
\label{LDPsum}
V_\mathrm{eff}^\mathrm{DP}(n_{\mathrm{max}})
=&\;\frac{|e\vec B|}{8\pi^2}\int_{-\infty}^\infty \!\!\!\! dp_z
\Big\{
\sum_\pm\sum_{n=0}^\infty|E_n^{\mathrm{DP}}|_{g_n=2}
\\ \nonumber
&\;
\qquad\;\;\;\;
+
\sum_\pm\sum_{n^\prime=1}^{n_{\mathrm{max}}} 
(|E_{n^\prime}^{\mathrm{DP}}|-|E_{n^\prime}^{\mathrm{DP}}|_{g_{n^\prime}=2})
\Big\}
\;.
\end{align}
The motivation for this truncation is that when the field-dependent self-energy corrections to the anomalous $g_n(B)-2$ contributions are taken into account, the leading Landau level ($n_{\mathrm{max}}=1$) contribution dominates {\it i.e.} $|g_1(B)-2|\gg |g_2(B)-2|$, Fig.1 of~\cite{Ferrer:2015wca}.

To assess the leading $n_{\mathrm{max}}=1$ contribution to $V_\mathrm{eff}^\mathrm{DP}$, Ferrer {\it et. al.\/} obtain a finite expression with the integration 
\begin{align}
\label{intSample0}
\int_{-\Lambda}^\Lambda \!\!\!\! dp_z \sqrt{p_z^2+x^2}
= \Lambda\sqrt{\Lambda^2+x^2} 
+\frac{x^2\!\!}2 \ln \frac{\sqrt{\Lambda^2+x^2} +\Lambda }{\sqrt{\Lambda^2+x^2} -\Lambda }
\;.
\end{align} 
which one can expand for $ x/\Lambda \to 0 $ to find 
\begin{align}
\label{intSample}
\mathrm{S_g}(\Lambda)\!\!\int_{-\Lambda}^\Lambda \!\!\!\!\!dp_z \sqrt{p_z^2+x^2}
= \frac{x^2\!\!}2 \left[1+\ln\frac{m^2}{x^2}\right]\!+\Lambda^2\!+\frac{x^2\!\!}2 \ln\frac{4\Lambda^2}{m^2} 
\;.
\end{align} 
We note that the cutoff $\Lambda$ defines the usual quadratically divergent zero-point energy that is removed by subtraction. Furthermore, $\Lambda$ regularizes the logarithmic divergence. However, even if each {\it individual} anomalous moment Landau level correction in~\req{LDPsum} is only logarithmically divergent, there is no assurance that the lowest Landau level anomaly contributions render an accurate result, nor that a worse divergence is hidden in the Landau-level summation. An indication that such challenging contributions arise in the DP action lies in the weak field expansion of~\req{LDPsum}, which contains odd in $B$ field contributions.

\subsection{Extending DP action of Ferrer {\it et. al.\/} }
\label{DPpartD}

Ferrer {\it et. al.\/} performing the Weisskopf summation of DP-Landau eigen-energies as a function of the $g-2$ anomaly present the DP-based effective action~\req{LDP} where any value of $g$ and thus anomaly $a=g-2$ can be inserted. The determination of the relevant value of $a$ is another mathematically distinct procedure. Instead, we can explore the form of effective action~\req{LDP} by treating $g$ as a parameter. 

A separation of higher order corrections into those associated with magnetic moment anomaly dependence of $V_\mathrm{eff}^\mathrm{DP}$, and the value of the anomaly is not visible in the two-loop HES proper time approach of Ritus~\cite{Ritus:1975cf}. This is so since the Landau level structure is not visible in the proper time evolution operator, and thus the dominant (Landau-level dependent) anomalous moment correction is not recognized.

In order to explore the renormalizability issues facing the DP based action as a function of parameter $g$, we next compute the higher order Landau level corrections to the $n_{\mathrm{max}}=1$ result of Ferrer {\it et. al.\/}, evaluating as a function of $g$ the truncated in anomalous moment correction summation~\req{LDPsum}. Taking the finite part of~\req{intSample} we evaluate~\req{LDPsum} for different $n_{\mathrm{max}}$ values, normalized to the $g=2$ HES result {\it i.e.\/} 
\begin{align}
\label{RDPgeneral}
R^\mathrm{DP}(g,n_{\mathrm{max}})
=&\;\frac{V_\mathrm{eff}^\mathrm{DP}}{V_\mathrm{eff}^\mathrm{DP}|_{g=2}}
\\ \nonumber
&\!\!\!\!\!\!\!\!\!\!\!\!\!\!\!\!\!\!\!=
1+\frac
{\int dp_z\sum_\pm\sum_{n^\prime=1}^{n_{\mathrm{max}}} 
(|E_{n^\prime}^{\mathrm{DP}}|-|E_{n^\prime}^{\mathrm{DP}}|_{g_{n^\prime}=2})}
{\int dp_z\sum_\pm\sum_{n=0}^\infty|E_n^{\mathrm{DP}}|_{g_n=2}}
\;.
\end{align}
Considering above with $g$ independent of Landau level given and fixed to order $\alpha$ QED value
\begin{align}
\label{gelectron}
g_e=2+\frac{\alpha}{\pi}+\mathcal{O}(\alpha^2)
\;,
\end{align}
we evaluate
\begin{align}
\label{dRDPgeneral}
\delta R^\mathrm{DP}(g,n_{\mathrm{max}}) =&\;
\frac{1-R^\mathrm{DP}(g,n_{\mathrm{max}})}{\alpha/2\pi}
\;.
\end{align}

In~\rf{fig:DPnmax} we plot $\delta R^\mathrm{DP}(g_e,n_{\mathrm{max}})$, \req{dRDPgeneral}, as a function of $n_{\mathrm{max}}$, choosing a fixed field strength $eB/m^2=100$. The key feature in \rf{fig:DPnmax} is that the relative strength of the correction to effective action by $\delta R^\mathrm{DP}(g,n_{\mathrm{max}})$ increases linearly as a function of $n_{\mathrm{max}}$. Each successive increase in $n_{\mathrm{max}}$ increases the anomalous contribution, diverging linearly with $n_{\mathrm{max}}\to\infty$, which superposes with the logarithmic divergence seen in \req{intSample}.

With $g$ fixed and independent of Landau level, an assumption inherent to the form of the action, the linear divergence recognized above suggests that the DP action is nonrenormalizable. The escape condition is as follows: Ferrer's~\cite{Ferrer:2015wca}'s evaluation in Ritus' framework for strong field QED shows that the effective $|g-2|$ decreases with Landau level. In the event that this decrease is found to be faster than linear, the effective action so derived could still be renormalizable.

On the other hand this conflicts with a further necessary condition to assure renormalizability: $|g-2|$ needs asymptotically to approach 0 in the large Landau level limit. Clearly this requirement cannot be met by most if not all particles including the electron, since $g\neq2$ from the start due to non-QED contributions, which give a fixed $|g-2|$ at all Landau levels. We thus conclude that the first order formulation of QED with $g\ne 2$ cannot be renormalized considering that no QED particle can be exactly at $g=g_\mathrm{D}$.

\begin{figure}[h]
\begin{center}
\includegraphics[width=0.68\columnwidth]{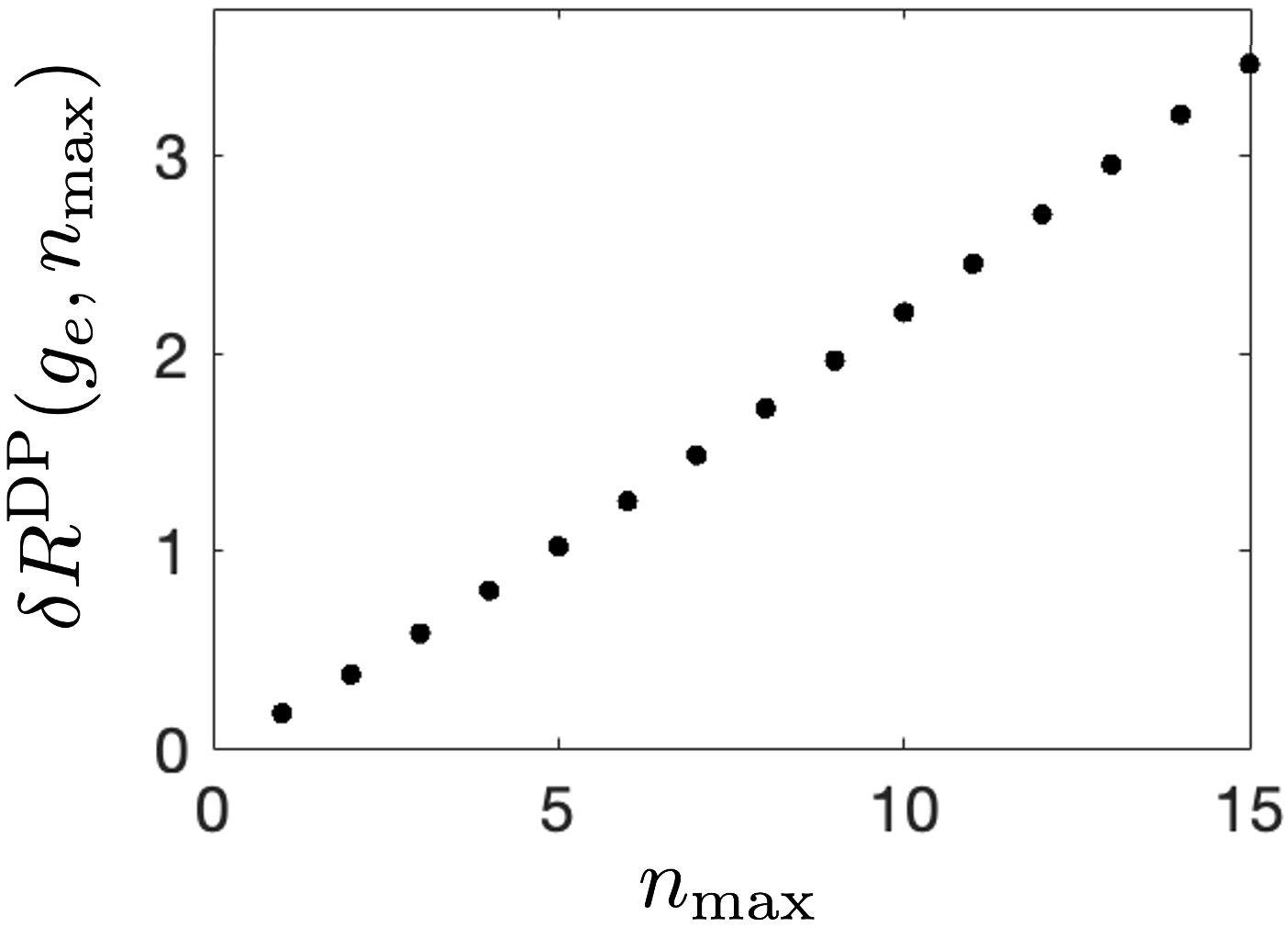}
\caption{The DP based anomalous contribution to the action~\req{dRDPgeneral}, plotted as a function of $n_{\mathrm{max}}$, at fixed $g=g_e$ (\req{gelectron}) and field strength $eB/m^2=100$. \label{fig:DPnmax}}
\end{center}
\end{figure}

Setting aside this problematic behavior related to the full summation of the Landau levels, the finite truncation by Ferrer {\it et. al.\/} allows us to further explore the $g$-dependent properties of the action. We consider the leading $n_{\mathrm{max}}=1$ result, Eq.\,(A.12) of~\cite{Ferrer:2015wca}, which we write explicitly as a function of $g$:
\begin{align}
\label{LDP}
V_\mathrm{eff}^\mathrm{DP}(n_{\mathrm{max}}\!=\!1)
=&\;\frac{\alpha B^2}{6\pi}\Big\{
\ln\Big(\frac{|e\vec B|}{m^2}\Big)
\\ \nonumber
&\!\!\!\!\!
- \frac{3|e\vec B|}{2 m^2}\Big(\frac {g_1}2-1\Big)^2\Big(\ln\Big(\frac{2|e\vec B|}{m^2}\Big)+2\Big)
\Big\}
\;.
\end{align}
The first term is the HES ($g=2$) expression, followed by the anomalous magnetic moment contribution: in~\cite{Ferrer:2015wca} the value of $g_1$ given by~\req{T1form} is applied, while we keep $g_1$ as a parameter. The relative minus sign between~\req{LDP} and Eq.\,(A.12) of~\cite{Ferrer:2015wca} distinguishes vacuum energy from the effective potential.

We display for three magnetic field backgrounds in~\rf{fig:DPFerrer} the DP action as presented by Ferrer {\it et. al.\/} as a function of $g$ and normalized to the $g=2$
\begin{align}
\label{RDP}
R^\mathrm{DP}(g,n_{\mathrm{max}}\!=\!1)
=&\;
1-
\frac{3|e\vec B|}{2 m^2}\Big(\frac {g_1}2-1\Big)^2
\frac{\ln\Big(\frac{2|e\vec B|}{m^2}\Big)+2}
{\ln\Big(\frac{|e\vec B|}{m^2}\Big)}
\;.
\end{align}
In this presentation \req{RDP} peaks at $g=2$, and becomes more sharply peaked for stronger $B$ fields. The behavior is akin to our KGP evaluation, compare~\rf{fig:Veffexpg}, where Lorentz scalar contributions to effective action (coefficient $f_{20}$) exhibit a smooth peak at $g=2$.

\begin{figure}[h]
\begin{center}
\includegraphics[width=0.95\columnwidth]{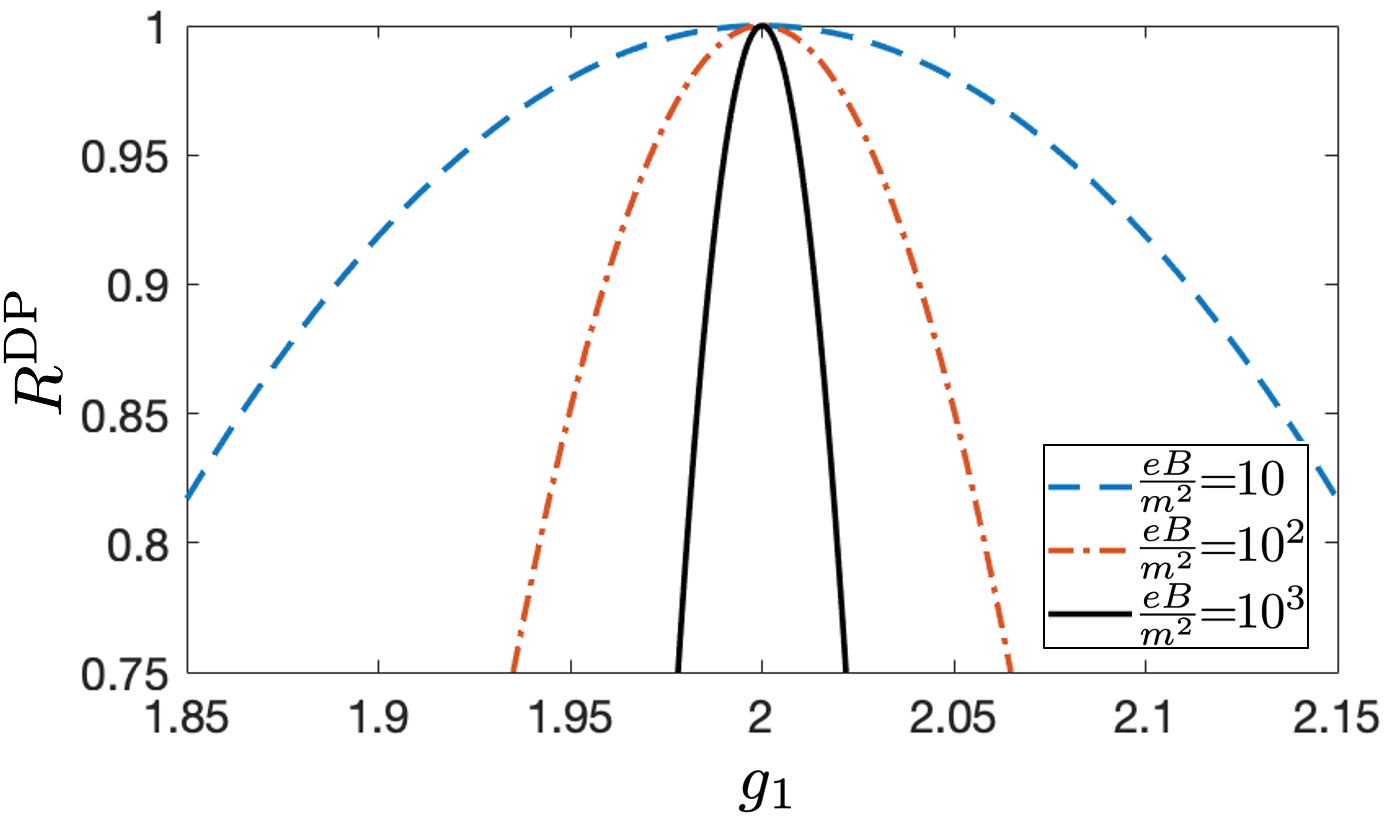}
\caption{Normalized at $g=2$ DP action, given by~\req{RDP}. \label{fig:DPFerrer}}
\end{center}
\end{figure}

\section{Discussion, Conclusions and Outlook} 

Difficulties of pQED as a stand-alone theory have been known for some time, beginning with the work of G. K\"all\'en~\cite{Kallen:1957ib, Kallen:1972pu}, and perturbative-pQED is believed by many to be semi-convergent only. Exploration of $g\ne 2$ in a renormalizable theory requires the dimension-4 $g$-QED based on the KGP equation. However, $g$-QED has to begin with 8 degrees of freedom and appropriate division into two half-Hilbert spaces is required. Restriction to the usual Dirac-like 4 degrees of freedom is difficult, as a theory with $g\ne 2$ is in general not unitary~\cite{Veltman:1997am}. 

We resolved this problem by proposing a new eigenstate sorting based on the sign of $K$, see~\req{spectrum1} in~\rs{periodicg}, leading to a self-adjoint theory that retains Poincar\'e symmetry and contains a complete set of particle-antiparticle states, and thus preserves probability in time evolution and analyticity as function of $g$, up to a countable set of singular points. A consequence of this solution is that the Dirac value $g=g_\mathrm{ D}=2$ is a cusp point of the effective action $V_{\mathrm{eff}}$,~\req{Veffab} evaluated in renormalizable $g$-QED approach. In~\rs{Bogsum} we described equivalence with the periodic in $g$ Bogoliubov coefficient summation method~\cite{Evans:2022fsu}. 
 
While~\req{VeffabF} is an analytic function of $g$, the integral of~\req{VeffabF} with the proper time weight~\req{Veffab} does not exist for $|g|>2$. Thus a naive extension of HES effective action to $|g|>2$ is not possible. This parallels the observation that the Klein-Gordon-Pauli operator~\req{EoMg} is not self-adjoint for $|g|>2$. We have shown how the eigenstate level crossing can be recognized and states assigned to half-spaces of the full Hilbert space, leading to a natural self-adjoint extension and a valid theoretical $g$-QED framework for $|g|>2$. The cusp and related nonperturbative in $g$ effects arise considering the self-adjoint extension described. The origin of the cusp singularity is in the periodic crossing of eigen energies in the spectrum of Landau eigenstates seen in upper section of~\rf{fig:Veffexpg} showing the quantity $K$, see~\req{spectrum1}.

We have shown cusps at $g=g_{\rm D}$ for two physical quantities computed for arbitrary $g$:\\
$\bullet$ The renormalization group coefficient $b_0$ proportional to function $f_{1,0}$, see~\rf{fig:Veffexpg};\\
$\bullet$ The light-by-light scattering in the long-wavelength limit comprising a smooth function $f_{2,0}$, and for the term $(\vec E\cdot\vec B)^2$ the cusp function $f_{0,2}$, see~\rf{fig:Veffexpg}. \\
We have checked that these results can be arrived at directly by the method of $\zeta$-function regularization following Weisskopf~\cite{Weisskopf}. Our results agree with earlier perturbative work in the fundamental domain $-2\le g\le 2$: the functional dependence on $g$ is explicit and the same for the vacuum polarization as had been obtained in Ref.\,\cite{AngelesMartinez:2011nt} in Eq.\,(56). We have shown by explicit computation that an expansion around $g=0$ is valid for $|g|\le 2$ only. 

We believe that our results imply that the pQED expansion around $g=g_\mathrm{D}$ is incomplete at sufficiently high order: Imagine that we partially resume $g-2$ diagrams with Dyson-Schwinger method, finding an effective electron with $g>2$. In the next step we want to compute the vacuum polarization inserts in other $g-2$ diagrams. Attempts in pQED framework will encounter new divergences as the $g-2$ correction is dimension-5 operator. On the other hand, we can accomplish this task in $g$-QED: we use the non-perturbative in $g$ renormalization group coefficient $b_0$ to characterize the vacuum polarization loop insert and there are no new divergences. However, the result contains the cusp, and thus is different from the finite order perturbative expansion of pQED.

We have shown in~\rs{DPpartB} that the cusp in $g$ appears in the truncated Dirac-Pauli approach based studies~\cite{Diet78, Ferrer:2015wca}. Dittrich~\cite{Diet78} omitted a term arising in the squared DP wave equation (\req{commutDP}). There is so far no argument known suggesting that this omission can be considered as a valid approximation to the DP solution in certain EM field configurations. However, it is notable that the remainder leads to a similar $g$-dependence of the effective actions as in the KGP approach. The singular cusp properties are the same, up to a field dependent $g$-anomaly proportional modification of particle mass. 

In~\rs{DPpartC} we explored the recent work of Ferrer {\it et. al.\/}~\cite{Ferrer:2015wca}, which evaluates the DP based action for pure magnetic fields, using the exact DP wave equation. In~\rs{DPpartD} we exploited the analytical properties of their Landau level summation procedure allowing for $g$ to be treated as a parameter. We demonstrated explicitly how the DP renormalizability issues surface from the exact solution: the summation is linearly divergent when a fixed anomalous moment correction enters all Landau levels,~\rf{fig:DPnmax}. Nonetheless, we believe that the truncation of Landau levels with anomalous moment carried out by Ferrer {\it et. al.\/} offers an encouraging development, as each individual anomalous contribution is only logarithmically divergent. The resulting $g$-dependence of the leading Landau level correction to the action in~\rf{fig:DPFerrer} exhibits a peak in the effective action at $g=2$, reminiscent in appearance to the KGP result~\rf{fig:Veffexpg}.

The works of Dittrich and Ferrer {\it et. al.\/} on DP based effective action are like our KGP approach --- in disagreement with a monotonically increasing analytical continuation from $|g| \le2$ to $|g|> 2$ as is usually presumed in QED. The results following from these DP approaches agree with the cusp at $g=2$ we described here. This supports the possibility that this singularity has a more general presence in QED, beyond the second order fermion KGP approach. Our findings are summarized in table~\ref{summarycusp}.
\begin{table}[h]
\begin{tabular}{|c|c|c|c|}
\hline
 & KGP & DP \\
\hline\hline
beta-function & 
\!\!\!\!\!\!\!\!\!\!\!cusp & 
cusp (Dittrich-DP~\cite{Diet78}) \\ 
\hline
HES action (pure $B$) & 
\!\!\!\!\!\!\!\!\!\!\!\!\! peak &
\!\!\!\!\!\!\! peak (Ferrer-DP~\cite{Ferrer:2015wca}) \\ 
 \hline
HES action (pure $E$) & 
 peak~\cite{Evans:2022fsu} &
 peak (Dittrich-DP~\cite{Diet78}) \\ 
 \hline
HES action ($E\cdot B$) & 
 cusp~\cite{Evans:2022EB} &
 cusp (Dittrich-DP~\cite{Diet78}) \\ 
\hline
\end{tabular}
\caption{Behavior at $g=2$ in the listed quantities: where cusps and smooth peaks occur.
\label{summarycusp}
}
\end{table}

There have been a lot of discussion and challenges since the cusp property has been recognized. However, we note that the presence of the cusp shouldn't be taken as a dramatic development because its behavior is similar to that of the essential singularity ($e^{-\pi m^2/e|E|}$) in the imaginary part of HES action. Both the essential singularity at zero electric field, and the singular point at $g=2$ for a pure electric field have been shown to be related to the constancy (infinite character) of the field. Both singular points go away when the finite extent of the external field is taken into account~\cite{Kim:2009pg, Evans:2022fsu}.

It is the remaining cusp in pseudoscalar $(\vec E\cdot\vec B)$ which is more profound and needs to be further explored as it may signal a novel singularity of the theory linked to a symmetry breaking term. To answer this question requires exploring how $(\vec E\cdot\vec B)$-dependent action behaves when taking into account the finite extent of EM fields. We will further explore the Bogoliubov summation method for finite spanning fields, using the solution by Kim, Lee and Yoon for a localized pseudoscalar profile: a Sauter-type $E$ field with a parallel constant $B$ field~\cite{Kim:2008yt}.

Our work has already served in Ref.\;\cite{Evans:2022EB} in an in depth exploration of its implications, including consideration of environments in which the nonperturbative cusp effect produces a notable difference from the perturbative in QED treatment of $g-2$. This difference appears to be most pronounced in magnetar environments of supercritical magnetic and subcritical electric fields~\cite{Ferrer:2015wca, Korwar:2017dio, Kim:2021kif, Kim:2022fkt}, where radiative corrections and mass catalysis effects become important~\cite{Miransky:2015ava}. For predictions of particle production in this domain, perturbative expansion in orders of $\alpha$ breaks down.

In summary, this analysis shows how a complete theory of a point-like fermion with $|g|>2$ can be constructed within $g$-QED in order to allow dynamical description of real world spin-1/2 particles. We have obtained the HES effective potential for an elementary particle with gyromagnetic ratio $g\ne 2$ nonperturbatively in $g$, see~\req{Veffab} and~\req{meroexpB}. We demonstrated a cusp-singularity as a function of $g$ at the Dirac value $g=g_\mathrm{ D}=2$. We have shown how this cusp enters the $\beta$-function and $(\vec E\cdot \vec B)^{2n}$ terms of light-light scattering. An interesting theoretical consequence is the possibility of asymptotic freedom in an Abelian theory with anomalous magnetic moment originating in the reversal in sign of the renormalization group coefficient $b_0$ for $g$ in specific domains much different from $g=2$.

\end{document}